# Extension of Ashby's performance indexes in mixed tension-bending solicitation

**O. Bouaziz**[*1,2], **J.P. Masse**[1],

[1]*ArcelorMittal Research, Voie Romaine-BP30320, 57283 Maizières-lès-Metz Cedex, France*
[2]*Centre des Matériaux, Ecole des Mines de Paris, CNRS UMR 7633, B.P. 87, 91003 Evry Cedex, France*

*Abstract*
Ashby's performance indexes are a fundamental tool for material selection especially for structures lightening. Unfortunately the indexes are available only for simple mechanical solicitation as pure bending or pure tension. For real applications, it is required to have a performance index for combined solicitations. This publication proposes an approach to develop this kind of extended performance index and shows the exploitation in order to compare materials performance in more realistic situations.

## 1. Introduction

In order to compare the performances of different structural materials, Ashby developed twenty years ago a key methodology based on performance indexes suitable to quantify the potential of lightening of structures [1-4]. Unfortunately, in the literature, the mechanical performances indexes are only define in pure bending or in pure tension solicitations as :

$$PI_t = \frac{E}{\rho} \quad (1)$$

for tension and

$$PI_f = \frac{E^{1/3}}{\rho} \quad (2)$$

for bending.

Unfortunately for real applications, it is required to have a performance index for combined solicitation. This publication proposes an approach to develop this extended performance index.

## 2. Extended performance index

In the case of combined bending and tension of a plate of length L as described in Fig1, the differential equation giving the deflection y is :

$$E.I\frac{d^2y}{dx^2} - F_2.y = -\frac{F_1.x}{2} \quad (3)$$

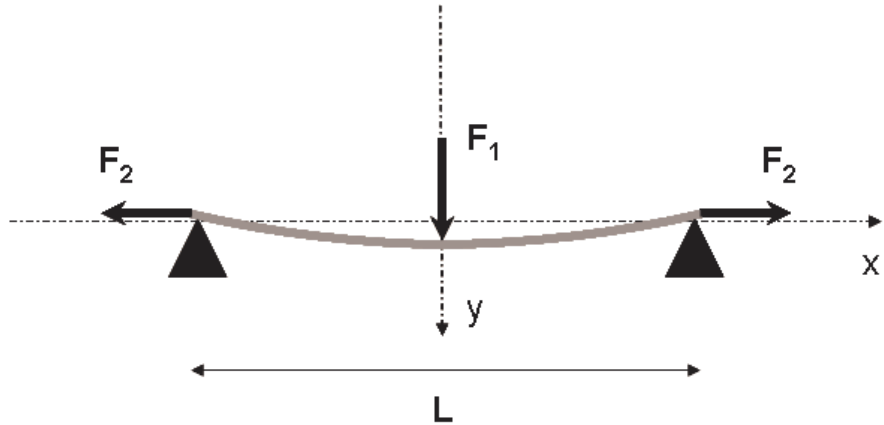

Figure 1 : Plate under combined bending and tensile forces.

The solution of this equation is the superposition of the particular solution $y_1$ respecting $\dfrac{d^2 y_1}{dx^2} = 0$ and $y_2$ the solution of the differential equation without the right end term :

$$E.I \dfrac{d^2 y_2}{dx^2} - F_2 . y_2 = 0 \quad (4)$$

So

$$y_1 = \dfrac{F_1 . x}{2.F_2} \quad (5)$$

And
$$y_2 = A.ch(k.x) + B.sh(k.x) \quad (6)$$
with
$$k = \sqrt{\dfrac{F_2}{E.I}} \quad (7)$$

A and B can be identified using the boundary conditions : $y(x=0) = 0$ and $\dfrac{dy}{dx}\left(x = \dfrac{L}{2}\right) = 0$

The first one imposed A=0. As :
$$\dfrac{dy}{dx} = B.k.ch(k.x) + \dfrac{F_1}{2.F_2} x \quad (8)$$
The second one imposed :
$$B = -\dfrac{F_1}{2.F_2} \cdot \dfrac{1}{k.ch\left(\dfrac{k.L}{2}\right)} \quad (9)$$

Finally :
$$y\left(x = \frac{L}{2}\right) = \frac{F_1}{2.k.F_2}\left(\frac{k.L}{2} - \tanh\left(\frac{k.L}{2}\right)\right) \quad (10)$$

It is interesting to notice that the power extension of tanh function gives :
$$\tanh\left(k.\frac{L}{2}\right) = \frac{k.L}{2} - \frac{1}{3}\left(\frac{k.L}{2}\right)^3 + \frac{2}{15}\left(\frac{k.L}{2}\right)^3 \quad (11)$$

$$y\left(x = \frac{L}{2}\right) = \frac{F_1}{F_2}\left(\frac{k^3.L^3}{E.I} - \frac{k^4.L^5}{60}\right) \quad (12)$$

$$y\left(x = \frac{L}{2}\right) = \frac{F_1.L^3}{48E.I}\left(1 - \frac{F_2.L^2}{10.E.I}\right) \quad (13)$$

Using Eq.13, it is known that the increase of rigidity is a linear function of the force $F_2$. In addition Eq10 indicates that the increase of rigidity increases as tanh function. So it is proposes to express the mixed performance index as :
$$PI_m = \frac{E^\alpha}{\rho} \quad (14)$$
With
$$\alpha = \frac{1}{3} + \frac{2}{3}.\tanh\left(\frac{F_2}{F_1}\right) \quad (15)$$

The evolution of $\alpha$ is drawn in Fig2 as a function of the ration between bending force and tensile force $\frac{F_2}{F_1}$. It is shown that for $\frac{F_2}{F_1} \geq 2$, it can be considered that condition of pure tension are achieved. In order to illustrate the approach, this expression is now applied to the case of comparison between steel and aluminium. In Figure 3, the ratio (E=210GPa, density:7.8) between the mixed performance index of steel divided by the mixed performance index of aluminium (E=70GPa, density:2.7) is plotted as a function of the ration between bending force and tensile force $\frac{F_2}{F_1}$.

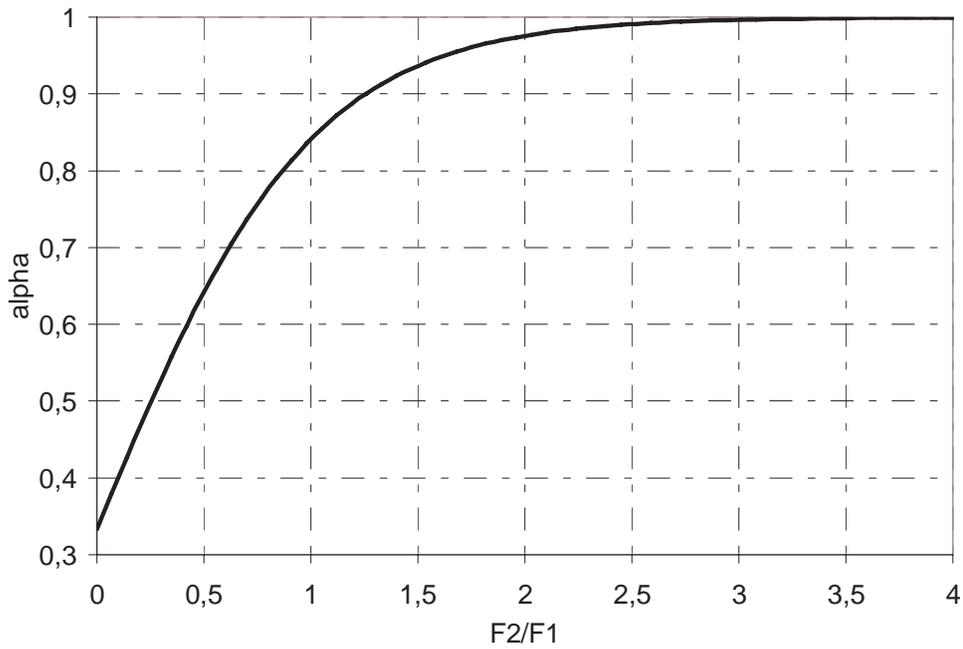

Figure 2 : Evolution of $\alpha$ for the mixed performance index as a function of the ratio between bending force and tensile force

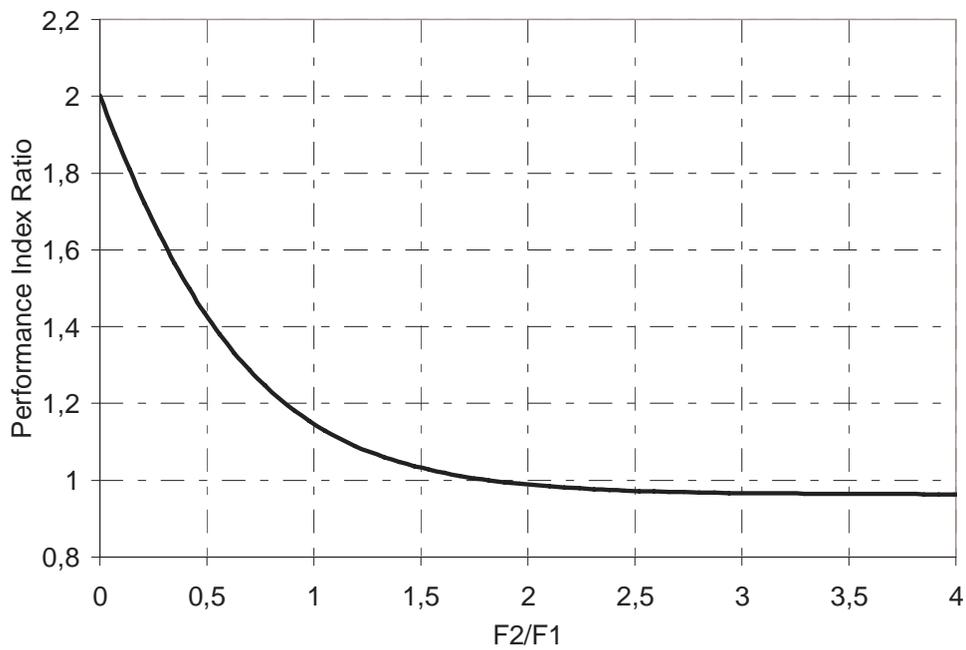

Figure 3 : Ratio between the mixed performance index of steel divided by the mixed performance index of aluminium as a function of the ration between bending force and tensile force.

## 3. Conclusions

For the assessment of the potential of a given material for structures lightening, Ashby's performance indexes are a fundamental tool for material selection. It has been noticed in this publication that these performance indexes are available only for simple mechanical solicitation as pure bending or pure tension. For real applications, it is required to have a performance index for combined solicitations. This publication proposed an approach to develop this extended performance index and shows the exploitation in order to compare materials performance in more realistic situations. The same kind of methodology can be performed for other mixed solicitation as tension-torsion or bending torsion for instance.